\documentstyle[prc,aps,epsf,preprint,tighten]{revtex}

\def\CF#1#2#3#4{#1 {\bf #2}, #4 (#3)}  

\def\etal {{\it et~al.}}
\def\fbs {Few-Body~Systems}
\def\fbss {Few-Body~Systems~Suppl.}
\def\hpa {Helv.~Phys.~Act.}
\def\plett {Phys.~Lett.~B}
\def\npa {Nucl.~Phys.~A}
\def\prev {Phys.~Rev.}

\def\prd {Phys.~Rev.~D}
\def\prc {Phys.~Rev.~C}
\def\sjnp {Sov.~J.~Nucl.~Phys.}

\def\ncim {Nuovo~Cim.}
\def\prl {Phys. Rev.~Lett.}
\def\epja {Eur.~Phys.~J.~A}
\begin{document}
\draft \preprint{ }
\title{
Relativistic Contributions to Deuteron Photodisintegration\\ in the
Bethe--Salpeter Formalism}
\author{K. Yu. Kazakov\thanks{\it E-mail: kazakovk@ifit.phys.dvgu.ru}
    and S. Eh. Shirmovsky\thanks{\it E-mail: suskovs@ifit.phys.dvgu.ru} }
\address{
Laboratory of Theoretical Nuclear Physics,\\ Far Eastern State
University,\\ Sukhanov Str. 8, Vladivostok, 690600, Russia}
\date{\today}
\maketitle
\begin{abstract}

In plane wave one-body approximation the reaction of deuteron
photodisintegration is considered in the framework of the Bethe--Salpeter
formalism for two-nucleon system. Results are obtained for deuteron vertex
function, which is the solution of the homogeneous Bethe--Salpeter
equation with a multi-rank separable interaction kernel, with a given
analytical form. A comparison is presented with predictions of
non-relativistic, quasipotential approaches and the equal time
approximation. It is shown that important contributions come from the
boost in the arguments of the initial state vertex function and the boost
on the relative energy in the one-particle propagator due to recoil.

\end{abstract}

\pacs{25.10+s; 25.20.Dc; 21.45+v; 11.10.St}

\section{Introduction}
\label{sec:intro}

Recently experiments on elastic and quasi-elastic lepton-nucleus
scattering have been considered as one of the fundamental and
reliable resources of probing structure of nuclei over a wide
range of energy-momentum transfer. A specific place among such
reactions is taken by deuteron photodisintegration. The reaction
$\gamma+d \rightarrow p+n$ has been thoroughly investigated in the
region of small and medium energies of the incoming
$\gamma$-quantum (an excellent review of experiments and
theoretical frameworks applied to the study of the process can be
found in Ref.~\cite{ArSan}). The theoretical analysis of
experimental data obtained in these kinematic regions produced an
important information on deuteron wave function, allowed one to
discriminate various contributions to the differential cross
section which stem from meson-exchange currents, isobar
configurations, spin-orbit and further relativistic corrections.

At the present time deuteron photodisintegration is rated among
the leading trends at experimental facilities around the world.
Here it is
worth-while to point out a performed experiment with the polarized
LADON gamma ray beam~\cite{AD98} to investigate the existence of
narrow dibaryonic resonances in reaction on nuclei at a low
excitation energies, an experimental program carried out on the
linearly polarized photon beam at YEREVAN Synchrotron to study the
cross section asymmetry of the deuteron photodisintegration
process in the energy range 0.9--1.7 GeV for proton center-of-mass
angle of 90$^\circ$~\cite{erevan}. This is connected the problem
of the validity of the constituent quark counting rules at
energies of a few GeV. Present experiments at SLAC with photons of
an energy 2.8 GeV~\cite{Belz} and proposed experimental programs
in RCNP Cyclotron with photon beam at an energy up to 8
GeV~\cite{RC} allow one to focus on investigation of hadronic
systems at quark level.

Future studies at the TJNAF are approved to extend measurements on
the differential cross section to a wide range of reaction angles
at high energies. The first measurements of the cross section from
$\gamma+ d\rightarrow p+n$ up to 4.0 GeV are in good agreement
with previous low energy measurements~\cite{TJ5}. A comparison of
the high precision Mainz data~\cite{Mainz,Mainz1} (over the photon
energy range 100--800 MeV) with meson-exchange models
incorporating relativistic effects~\cite{wilhelm,laget,jaus} shows
that one still lacks of the ultimate conclusion on the role of
non-nucleonic degrees of freedom in nuclei. It is also understood
that further development of deuteron photodisintegration theory is
needed and consistent treatment of relativistic effects have to be
applied. This demands a construction of a genuine relativistic
formalism for the description both the deuteron structure and the
reaction mechanism.

The formulation of a completely relativistic formalism of hadronic
bound states and reaction with them can be developed on the basis
field-theoretical Bethe--Salpeter (BS) equation for the
nucleon-nucleon (NN) scattering~\cite{RT1}. However approximate
methods evolved from the BS formalism due to substantial
mathematical and computational problems. We bear in mind the
quasipotential (QP) approach, which reduces the 4-dimensional BS
equation to a relativistic 3-dimensional equation: the
Logunov--Tavkhelidze~\cite{LT}, Blankenbeckler--Sugar
(BbS)~\cite{BbS} and Gross~\cite{gross} and other approximations.

Actually the relativistic description of the reaction with the
deuteron is extensively developed in the framework of the BS
formalism, which is explicitly Lorentz covariant, provides
two-body unitarity and includes nucleon and anti-nucleon degrees
of freedom in a hadronic state in a symmetrical manner. So far
applicability of the BS equation to reactions with the deuteron
has been bound to: elastic electron-deuteron
scattering~\cite{RT1}, elastic $pd$-backward
scattering~\cite{RT7}, inclusive quasi-elastic electron-deuteron
scattering~\cite{UmnK,ciofi} and description of the static
properties of the deuteron~\cite{HonzIsh,KapKaz}. The common
feature of these processes is that the reaction amplitude in the
impulse approximation is proportional to the averaging of current
operator between deuteron states. On the contrary, the
reaction amplitude for deuteron photodisintegration is a
non-diagonal matrix element between the incoming deuteron and
outgoing 2N state.

An approach towards a covariant description of the reaction $\gamma+ d\to
p+n$, in which the basic degrees of freedom are taken to be hadronic, is
developed in the framework of the dispersion relation technique for
laboratory photon energies $E_\gamma<400$~MeV~\cite{anisovich} (this
technique is appropriate for the analysis of partial wave amplitudes and
takes into account final state rescatterings). Other approach describes
deuteron photodisintegration as a simple parameterization of covariant
deuteron in terms of a hard component and imposing gauge invariance on the
cross section~\cite{nagor}. This approach looks into the onset of scaling
in exclusive photodisintegration for high energies in the range 1-4 GeV,
as it can be brought about by mechanisms which are different from those of
pQCD.

The BS formalism presents a separate view on the problem.  Formal application
of the BS formalism to deuteron electro- and photodisintegration can be found
in Ref.~\cite{KorSheb}. Here the rigorous derivation of the scattering
amplitude in terms of BS amplitude of the initial and final 2N states and
Mandelstam electromagnetic (EM) vertex, which comprises the one-body and
two-body parts, is proposed. Although the numerical analysis has not been
performed. In paper~\cite{hummel} it is proposed a framework based on the BS
equation approach. This is applied to both elastic and inelastic electron
scattering.  But complete calculations are performed within a QP framework.

The aim of this paper is to apply the fully relativistic analysis
of deuteron photodisintegration in the framework BS formalism, to
segregate and estimate contribution of various relativistic
effects in the differential cross section.

This paper is organized as follows. In Sec.~\ref{sec:formalism} we briefly
discuss a connection between the BS formalism and QP approach, and equal time
(ET) approximation. All the relativistic formulations of the 2N dynamics,
exploiting relativistic separable interactions, are applied to the study of
the deuteron and its inelastic observables. Basic formulae for definition of
the Minkowski-space BS amplitude for bound and 2N continuum states are given.
We also introduce formulae for the relativistic separable interaction kernel.
Solving the BS equation with the separable interaction, we find the vertex
function used in computation of the unpolarized cross section.

In Sec.~\ref{sec:cross-section} we describe the procedure to derive deuteron
photodisintegration cross section. The EM interaction with 2N system in the
framework of the BS formalism is determined by the Mandelstam vertex, which
generally depends on properties of the interaction kernel. The scheme
incorporates two-body part of the Mandelstam vertex in order to guarantee the
gauge independence.

Sec.~\ref{sec:pwia} deals with derivation of the simplest contribution to the
EM current matrix elements of deuteron break-up, namely the plane wave
one-body approximation (PWOA). We discuss the transformation properties of
the deuteron vertex function between the laboratory and c.~m. frames.
Generally one-body part of the Mandelstam vertex involves half off-mass-shell
$\gamma$NN form factors. However, as consequence of gauge invariance the real
photon scattering amplitude does not contain off-shell effects. In the PWOA
the EM current matrix elements are proportional to deuteron vertex function
which is taken at certain value of the relative energy (relative time) and
3-momentum. These are directly related to the photon energy-momentum
transfer. Finally, we write down the expression for the differential
photo-disintegration cross section.

Sec.~\ref{sec:analysis} is devoted to analysis of relativistic effects
required by the principles of relativity. These are relativistic
kinematics and dynamics, the relative energy dependence (or retardation),
the Lorentz contraction and spin precession. We compare results of our
fully relativistic analysis with those of conventional non-relativistic
(NR) models, the QP approach and the ET approximation.

In Sec.~\ref{sec:remarks} we conclude about our results.

\section{Relativistic description of two-nucleon system}
\label{sec:formalism}

Formulation of integral equations for amplitudes is commonly taken as the
starting point for discussing relativistic scattering and bound state
problems in the strongly interacting systems within quantum field theory.

The off-shell $T$-matrix for the elastic scattering of two nucleons with the
relative 4-momentum $p$, $p^\prime$ and the total momentum $P$ satisfies to
the inhomogeneous BS equation. In the momentum space this is the
four-dimensional (4-D) integral equation with respect to the relative
momentum $k=(k_0,
{\bf k})$\footnote{For sake of simplicity we omit writing spinor, $\rho$-spin
and polarization quantum numbers of amplitudes below}
\begin{equation}
T(p,p^\prime;P)={\cal V}(p,p^\prime;P)+\frac{i}{4\pi^3}\int\!\! d^4 k \,
{\cal V}(p,k;P)G_0(k;P)T(k,p^\prime;P),\label{T-matrix-eqn}
\end{equation}

\noindent where ${\cal V}$ is an interaction kernel obtained by summing of
all irreducible 2N Feynman diagrams in a given field-theoretical model of the
NN interaction, and $G_0(k;P)$ --- the free two-nucleon propagator
\begin{eqnarray}
G_0(k;P)=\frac{[\hat P/2 + \hat k + m]^{(1)}}
{(\frac{P}{2}+k)^2-m^2+i\epsilon}
\cdot\frac{[\hat P/2 - \hat k + m]^{(2)}}
{(\frac{P}{2}-k)^2-m^2+i\epsilon}.
\label{propagator}
\end{eqnarray}

The BS amplitude for the 2N scattering state, $P^2=s>4m^2$, is expressed in
terms of the half off-shell $T$-matrix and the propagator function $G_0$ as
follows
\begin{eqnarray}
\chi(k;\hat p P)=\large[
4\pi^3i\delta^{(4)}(\hat p-k)
-G_0(k;P)T(k,\hat p;P)\large]\chi^{(0)}(\hat p;P),\qquad
\hat p\cdot P=0.\label{BS-cont}
\end{eqnarray}

\noindent where $\hat p$ denotes on-mass-shell relative 4-momentum
and $\chi^{(0)}(\hat p;P)$ is the amplitude for the motion of free
nucleons. The second term comprises rescattering contributions.

When the two-body system has a bound state of mass $M_d$, the $T$-matrix has
a pole at $P^2=M_d^2$
\begin{eqnarray}
T(p,p^\prime;P)\propto \frac{\Gamma(p;P)\bar\Gamma(p^\prime;P)}{P^2-M_d^2},
\label{pole}\end{eqnarray}

\noindent where $\Gamma$ is the vertex function and $\bar \Gamma$
is its conjugate. The vertex function of the 2N bound state
satisfies to the homogeneous BS equation with the same interaction
kernel
\begin{equation}
\Gamma(p;P)=\frac{i}{4\pi^3}\int\!\! d^4 k\,
{\cal V}(p,k;P)\psi(k;P),\qquad P^2=M_d^2,\label{vertex-eqn}
\end{equation}

\noindent where the bound amplitude is defined as:
\begin{equation}
\psi(k;P)=G_0(k;P)\Gamma(k;P).\label{BS-disc}
\end{equation}

The Eqns.~(\ref{T-matrix-eqn}) and~(\ref{vertex-eqn}) are
manifestly Lorentz covariant and preserve the two-body elastic
unitarity. Moreover both equations do not discriminate between the
positive and negative energy states, yielding transformation
properties of the calculated amplitudes consistent with the charge
conjugation and time invariance.

Although the BS calculations are feasible, a rigorous treatment of
Eqs.~(\ref{T-matrix-eqn}) and (\ref{vertex-eqn}) is rather
complicated due to the appearance of a relative energy in loop
integrals and the presence of strong singularities in the
interaction kernel. The great theoretical effort have been applied
to obtain 3-D bound-state equations from the 4-D one. A simple
method to obtain an approximate 3-D equation has been developed by
the QP approach, in which the BS equation is reduced to a 3-D
equation by making use of a new two-nucleon propagator with the
internal relative energy variable restricted to a fixed value.

But there are some shortcomings in the relativistic equations
obtained from the BS equation via the 3-D reduction. First of all
one meets conceptual difficulties with the consistent treatment of
both the 2N system and its EM interactions~\cite{wallace}.
Secondly by putting particles on mass-shell or using the positive
energy projection operators, one results in a equation which
violate the charge conjugation and CPT symmetries. Importance of
the constraint put by discrete symmetries is discussed in
Ref.~\cite{pascalutsa}.

An alternative choice, which have been made to study the EM
interactions, is the 'instant' or equal time
approximation~\cite{hummel,wallace1}. The instant-ET approximation
to EM current operator is a 3-D reduction consistent with charge
conjugation and unitarity. This approach has been applied to study
of the elastic electron scattering case and  the deuteron breakup
in electrodisintegration process.

In this paper we extend the ET choice in a systematic way to
describe the process of deuteron photodisintegration. In the PWOA
the ET approximation can be obtained simply by replacing the
initial deuteron vertex function by the BbS vertex function. The
relative energy variable in the one-particle propagator is
prescribed by the condition that the final state describes
on-mass-shell particles~\cite{hummel}.

The covariant BbS prescription, $P\cdot \hat k=0$, puts the relative energy
equal to
\begin{eqnarray}
\hat k_0 = \frac{1}{2P_0}(E_{\frac12{\bf P}+{\bf k}}^2-
E_{\frac12{\bf P}-{\bf k}}^2),
\nonumber\end{eqnarray}

\noindent where $E_{\bf k}=\sqrt{m^2+{\bf k}^2}$ and $\hat k$
denotes the restricted 4-vector $k$. The prescription leads to a
relativistic equation of motion for the moving 2N system. By means
of the boost it is transformed to the Schr\"{o}dinger-type
equation in the rest frame\footnote{A detailed and systematic
exposition of the covariant QP formalism for description of the
electromagnetic (EM) properties and reactions involving the
deuteron is given in Ref.~\cite{jaus-woolock}.}.

The BbS prescription is simple only in the 2N rest frame,
$P_{(0)}=(\sqrt{s},{\bf 0})$, where the BbS propagator
$G_{\text{BbS}}(\hat k;P)$ is given by
\begin{eqnarray}
G_{\text{BbS}}(\hat k;P_{(0)})=-2\pi i\delta(\hat k_0)
{\cal G}(\hat k;P_{(0)})\Lambda^{(1)}({\bf k}_1) \Lambda^{(2)}({\bf k}_2),
\label{propagator-BBS}\end{eqnarray}

\noindent where $\Lambda^{(i)}({\bf k}_i)$ is the positive energy
 projection
operators  and ${\cal G}(\hat k;P_{(0)})=\frac{4m^2}{E_{{\bf k}}(s-4E_{{\bf
k}}^2+i\varepsilon)}$. The delta-function in
Eq.~(\ref{propagator-BBS}) sets the relative energy equal to zero
putting both nucleons equally off-mass-shell. It is
an attractive feature in case of the deuteron because it treats
both nucleons in a symmetrical way and, as consequence, it is
consistent with the Pauli principle.

The QP wave function is defined in terms of the bound state vertex
 function
$\hat\Gamma(\hat k;P)$ defined in terms of the positive energy Dirac
spinors~\cite{jaus}:
\begin{eqnarray}
\phi_{\text{QP}}(\hat k;P)=
\sqrt{\frac{Q(\hat k;P)}{m}}{\cal G}(\hat k;P)
\hat\Gamma(\hat k;P),
\label{BbS-wf}\end{eqnarray}

\noindent where $Q(\hat k;P)=\case{P_0}{M_d}\sqrt{m^2-{\hat k}^2}$. The
equation satisfied by the vertex function follows from the BbS reduction of
the BS equation:
\begin{eqnarray}
\hat\Gamma(\hat p;P)=
\frac{1}{2\pi^2}\int\!\! d^3k\hat{\cal V}(\hat p,\hat k;P){\cal G}(\hat k;P)
\hat\Gamma(\hat k;P),
\label{BbS-eqn}\end{eqnarray}

\noindent where $\hat{\cal V}(\hat p,\hat k;P)$ is a Lorentz invariant
quasipotential with all relative 4-momenta are restricted by the BbS
condition. The QP wave function $\phi_{\text{QP}}({\bf k};P_{(0)})$ at the
rest frame of the deuteron, $P_{(0)}=(M_d,{\bf
 0})$, is expressed in terms
of Lorentz invariant
 $\sqrt{\case{P_0}{M_d}}\phi_{\text{QP}}(\hat k;P)$.  By
means of the boost relativistic equation of motion~(\ref{BbS-eqn})
for the moving deuteron transforms into equation in the rest
frame. This invariance property yields the NR equation for the wave function:

\begin{eqnarray}
\frac{M_d^2-4E_{\bf p}^2}{4m}\phi_{\text{QP}}({\bf p}_{(0)};P_{(0)})=
\frac{1}{2\pi^2}\int\!\! d^3k V({\bf p}_{(0)}, {\bf k}_{(0)}; P_{(0)})
\phi_{\text{QP}}({\bf k}_{(0)};P_{(0)}),
\label{quasi}\end{eqnarray}

\noindent where 3-momenta ${\bf p}$ and ${\bf k}$ in the moving
frame are mapped by the boost transformation to ${\bf p}_{(0)}$
and ${\bf k}_{(0)}$ at the rest frame, respectively, and $V$ is
the quasipotential  modified by `minimal relativity'.

Further the discussion concerns partial decomposition of the BS
vertex function of the deuteron $\Gamma(k;P)$. For definiteness at
the rest frame one has (we highlight dependence on the spin
projection):
\begin{equation}
\Gamma_{M}(p;P_{(0)})=\sum\limits_{\alpha} g_{\alpha}(p_0,|{\bf
p}|;\sqrt{s}) {\Gamma}_M^\alpha(-{\bf p}),\qquad (M=\pm
1,0),\qquad \sqrt{s}=M_d.\label{vert} \label{vertex-partial}
\end{equation}

\noindent Here summation index $\alpha$ is determined by the
following quantum numbers: $S=0,1$ --- spin, $L=0,1,2$ --- angular
momentum, $J=1$ -- total angular momentum and $\rho$-spin --- the
projection of the total energy spin of the nucleon and
anti-nucleon states; ${\Gamma}_M^\alpha(-{\bf p})$ is the
spin-angular functions and $g_{\alpha}$ is the partial amplitudes.
Eight partial states contribute to Eq.~(\ref{vertex-partial}). Apart
from two channels with the positive energy intermediate states,
viz. $^3S_1^{+}$--$^3D_1^{+}$, six 'extra' states, which account
for anti-nucleon degrees of freedom, come into play. In the
spectroscopic notations ${}^{2S+1}L_J^{\rho}$ these are

$${}^3P_1^{e}, {}^3P_1^{o}, {}^1P_1^{e}, {}^1P_1^{o}, {}^3S_1^{-}, {}^3D_1^{-},$$

\noindent where indexes $e$ and $o$ stands for the even and odd
parity relative to the $\rho$-spin functions. The partial
amplitudes ${}^1P_1^{e}$, ${}^3P_1^{o}$ are even and
${}^1P_1^{o}$, ${}^3P_1^{e}$ are odd functions in the relative
energy variable.

Influence of admixtures of $P$-states and their contribution in
interference with the positive energy states to observables in
deuteron breakup and elastic proton-deuteron backward scattering
calculated within the BS formalism is considered in
Refs.~\cite{RT7,BondBur,semikh}. It is shown that $N\bar N$ pair
EM current term in the NR approach can be constructed from the
$P$-state in the deuteron BS amplitude. The separate effect of the
pair current term, with respect to the one-body, is expected to
give constructive interference followed by a destructive
interference with increasing photon energy in the range 0.1--0.7~
GeV~\cite{chemtob}.

In the present paper we focus on the positive energy states only.
So we have two channels, $^3S_1^+-{}^3D_1^+$, and the corresponding
vertex functions can br written in the matrix form~\cite{KapKaz}:
\begin{eqnarray}
\sqrt{8\pi}\,\Gamma_{\lambda}^{^3S_1^{++}}(p;P_{(0)})&=& {\cal N}^2_{{\bf
p}}(m+\widehat{p}_{1})(1+\gamma_{0}) \widehat{e}_{\lambda}
(m-\widehat{p}_{2})g_0(p_{0},|{\bf p}|;s),\label{3s1}\\
\sqrt{16\pi}\,\Gamma_{\lambda}^{^3D_1^{++}}(p;P_{(0)})&=&-{\cal N}^2_{{\bf
p}}(m + \widehat{p}_{1})\nonumber\\ &\times& (1 +
\gamma_{0})\left(\widehat{\varepsilon}_{\lambda} +
\frac{3}{2}(\widehat{p_1}-\widehat{p_2}) \frac{(p\cdot e_\lambda)}{{\bf
p}^{2}}\right) (m-\widehat{p}_{2})g_2(p_{0},|{\bf p}|;s),\nonumber
\end{eqnarray}

\noindent where $s=M_d^2$, $p_1=(E_{{\bf p}},{\bf p})$ ¨
$p_2=(E_{{\bf p}}, -{\bf p})$ are on-mass-shell 4-momenta, ${\cal
N}^{-1}_{{\bf p}}=\sqrt{2E_{\bf p}(m+E_{\bf p})}$ is the
normalization factor and  $e_\lambda=(0,{\bf e}_\lambda)$ is 4-polarization
vector of the deuteron:
\begin{eqnarray}
\sum\limits_{\lambda=-1}^{+1} e_\lambda^\mu e_\lambda^{\nu
*}=-g^{\mu\nu}+\frac{P^\mu P^\nu}{M_d^2}, \qquad e_\lambda\cdot P=0.
\label{sum-e}\end{eqnarray}

We are interested in calculating  matrix elements of the EM
current operator between a state containing free nucleons and 2N
bound state. To this end we need to take into account change in
the state amplitudes, when the BbS prescription is applied to the
S-matrix element for this process. We can utilize the
normalization condition for the bound state amplitude. Let us
suppose that the kernel ${\cal V}$ is independent of the total
4-momentum $P$. Then we obtain
\begin{equation}
1=-\int \frac{d^4k}{2\pi^2i}\,\bar\Gamma(p;P)\left.
\frac{\partial G(p;P)}{\partial P^2}\right|_{P^2=M_d^2} \Gamma(p;P).
\end{equation}

\noindent In terms of the radial partial vertex functions~$g_L$
this conditions takes the form
\begin{eqnarray}
\frac{1}{2\pi^2 i M_d}\int\limits_0^\infty d k_0
\int\limits_0^\infty d|{\bf k}|\,{\bf k}^2
\frac{g_L(k_0,|{\bf k}|;s)^2
(E_{\bf k}-\frac{M_d}{2})}
{((\frac{M_d}{2}-E_{\bf k}+i\epsilon)^2-k_0^2)^2}=P_L,\qquad P_0+P_2=1.
\label{BS-norm}\end{eqnarray}

\noindent We find respectively for the QP vertex function ${\hat
g}_L$:
\begin{eqnarray}
\frac{2m^2}{\pi^2 M_d}\int\limits_0^\infty d|{\bf k}|\,{\bf k}^2
\frac{{\hat g}_L(0,|{\bf k}|;s)^2}{E_{\bf k}(M_d^2-4E^2_{\bf k})^2}=P_L.
\label{BBS-norm}\end{eqnarray}

\noindent One can deduce out of these Eqns. that the BS and QP
vertex functions are formally related to each other as follows
\begin{eqnarray}
\biggl.\frac{g_L(k_0,|{\bf k}|;s)}{E_{\bf k}-\frac{M_d}{2}}
\biggr|_{k_0=\frac{M_d}{2}-E_{\bf k}}\propto\sqrt{\frac{4
m^2}{\pi E_{\bf k}}} \frac{{\hat g}_L(0,|{\bf k}|;s)} {4E_{\bf
k}^2-M_d^2}. \label{BS-to-BBS}
\end{eqnarray}

\noindent In Eq.~(\ref{BS-to-BBS}) the QP function~(\ref{quasi}) is
related to the Schr\"odinger wave function by minimal relativity:
\begin{eqnarray}
\phi_{\text{QP}}({\bf
k};P_{(0)})\equiv\sqrt{\frac{m}{E_{\bf k}}}\phi_{\text{NR}}({\bf k}).
\nonumber\end{eqnarray}

The presence of strong singularities in the interaction kernel can
be avoided by special prescriptions referred to imply analytical
properties. The ladder approximation with the interaction kernel
of the BS equation ${\cal V}$ to be a sum of one boson exchange
diagrams is widely used in solving BS equation for the NN
scattering. In this case a solution of Eqn.~(\ref{vertex-eqn})
(for example, we refer to papers~\cite{RT7}-\cite{KapKaz}), can be
found in Euclidean space after a Wick rotation. However that
presents a substantial obstacle when one calculates observable in
terms of the BS amplitude. Actually an analytical continuation of
the solution in the complex $k_0$-plane back to the real axis is a
procedure which carries ambiguities and extremely laborious.  The
physical solution can be obtained via the method based on the
Perturbation Theoretic Integral Representation of
Nakanishi~\cite{nakanishi}. The BS equation for bound states is
solved in terms of a generalized spectral representation directly
in Minkowski space~\cite{kusaka}. But the approach is developed
for bound states in scalar theories.

An alternative way to solve the BS equation is to use a non-local separable
interaction kernel~\cite{RT4,RT5,schwarz}:
\begin{eqnarray}
{\cal V}(p,p^\prime;P)=
\sum_{a,b=1}^{N}
\lambda_{ab} v_a(p;P) v_b(p^\prime;P),
\label{sep-kernel}\end{eqnarray}

\noindent where $\lambda_{ab}$ is a symmetrical matrix. In this
case the bound state vertex function and the scattering $T$-matrix
is obtained in Minkowski space. They exhibit analytical properties
determined by form factors $v_a(p;P)$. The covariant form factors
includes a dependence on $P^2$, $p^2$ and $p\cdot P$. Leaving the
dependence on $p^2$ only means a simple procedure to construct
relativistic separable interactions to be used in the BS equation.
That is done in study of 2N states in the BS formalism with the
separable form of interaction, see Refs.~\cite{RT5,RT2,RT8}. These
form factors are not expected to be genuine separable
approximations to a realistic NN interactions.

In this paper we use the separable kernel of rank three ($N=3$ in
Eqn.~(\ref{sep-kernel})) for computation of the deuteron
photodisintegration cross section. This interaction kernel is a
relativistically covariant generalization of the NR Graz-II potential for
the description of the phase shifts of the NN scattering in the coupled
$^3S_1$--$^3D_1$ waves (details can be found in~\cite{RT3}). The
analytical properties of the radial vertex function is determined by poles
in the relative energy:
\begin{eqnarray}
g_0(k_0,|{\bf k}|;s)=A(s)\frac{1-\gamma_1
k^2}{(k^2-\beta_{11}^2)^2}
+B(s)\frac{k^2}{(k^2-\beta_{12}^2)^2},\nonumber\\ g_2(k_0,|{\bf
k}|;s)=C(s)\frac{k^2(1-\gamma_2k^2)}{(k^2-\beta_{21}^2)^2
(k^2-\beta_{22}^2)^2}, \quad
s=M_d^2,\label{BS-vertex}\end{eqnarray}

\noindent where $k^2=k_0^2-{\bf k}^2$, the coefficients $A,B$ ¨
$C$ are determined by the homogeneous set of three algebraic
equations, the parameters $\beta_{ab}$ and $\gamma_a$ are chosen
to reproduce $^3S_1$--$^3D_1$ NN scattering phase shifts up to a
laboratory energy of $E_{\text{Lab}}=500$~MeV, the low-energy NN
scattering parameters and the static deuteron properties (the
binding energy, the quadrupole and magnetic moments). We adopted
the parameters of the interaction kernel corresponding to the
value of ${}^3D_1^{+}$-state probability $P_2=4$~\% and 6~\%. The
actual parameters $\beta_{ab}$ and $\gamma_a$ are chosen to have
such values that the resulting $T$-matrix $T_{LL^\prime}(p_0,{\bf
p},p_0^\prime,{\bf p}^\prime;s)$ satisfies the exact two-body
unitarity relation at least up to a nucleon kinetic energies
$E_{\text{Lab}}=
\case{2}{m}\beta_{11}\left(m+\case{\beta_{11}}{4}\right)$ with
$\beta_{11}=231$~MeV, which is about 500 MeV in the laboratory
system.

This is repeated using the QP approximation to the BS equation.
The QP vertex function is the solution of the homogeneous BS
equation with the BbS version of the Green
function~(\ref{propagator-BBS}) and the interaction kernel
equivalent to the NR Graz-II potential. The values of the
parameters $\lambda_{ab}$ are changed relative to the BS
parameters to reproduce resulting deuteron properties, low-energy
scattering parameters and the phase shifts. The resulting radial
part of the deuteron vertex function ${\tilde g}_L$ depends on the
relative 3-momentum as follows
\begin{eqnarray}
\hat g_0(0,|{\bf k}|;s)=\hat A(s)\frac{1+\gamma_1 {\bf k}^2}
{({\bf k}^2+\beta_{11}^2)^2}
+\hat B(s)\frac{{\bf k}^2}{({\bf k}^2+\beta_{12}^2)^2},\nonumber\\
\hat g_2(0,|{\bf k}|;s)=\hat C(s)\frac{{\bf k}^2(1+\gamma_2{\bf k}^2)}
{({\bf k}^2+\beta_{21}^2)^2({\bf k}^2+\beta_{22}^2)^2},
\quad s=M_d^2.\label{QP-vertex}\end{eqnarray}

The particular feature of the BbS reduction is that all propagators are
reduced to the static form. Speaking in language of the meson-exchange
model, the BbS reduction completely ignores the retardation or the
relative energy dependence in the BS amplitude which arises from
non-instantaneous effects in the NN interaction.

\section{Deuteron photodisintegration cross section}
\label{sec:cross-section}

Let us consider disintegration of a deuteron with total 4-momentum $K$ by a
photon
 with 4-momentum $q^{\mu}$, $q^2=0$, into a free neutron-proton (np)
pair,
 characterized with the total and relative 4-momenta $P$ and $p$,
respectively. In the rest frame of the np pair, i.~e.
$P_{(0)}^{\mu}=(\sqrt{s},{\bf 0})$, $p^{\mu}=(0,{\bf p})$, where $\sqrt s$
--- is
 the total energy of the pair, the differential absorption cross
section of a
 photon with energy $\omega$ can be written as

\begin{equation}
\frac{d\sigma}{d\Omega_p}=\frac{\alpha}{16\pi s} \frac{\mid {\bf
p}\mid}{\omega} \overline{\mid \varepsilon_\lambda\cdot {\cal
M}_{\text{fi}} \mid^{2}} \label{CMC}\end{equation}

\noindent with $\alpha=e^2/(4\pi)$ is the fine structure constant,
${\cal M}_{fi}^\mu$ --- the invariant amplitude, which is the
transition matrix element ${\cal M}_{fi}^\mu= \langle f \mid
\widehat{J^\mu}\mid i\rangle$ of the EM current operator
$\widehat{J^\mu}$ between the deuteron bound state and the 2N
continuum; $\varepsilon_\lambda^\mu$ is a photon polarization
4-vector with $\lambda=\pm 1$. Momentum conservation at the
photon-deuteron vertex gives $K+q=P$.

Since polarizations of the particles involved in the process will
 not be
considered here, averaging and summing over the photon and
 nuclear
polarizations in the initial and final states,
 respectively, are assumed.
We may choose such a coordinate system
 where the photon 3-momentum is along
a $Z$ axis:
 $q^\mu=(\omega,0,0,\omega)$. In the laboratory system, being
the
 rest frame of the deuteron, the deuteron 4-momentum $K_{(0)}=(M_d,{\bf
0})$ and the photon energy is denoted as $E_\gamma$.

In experiments on two-body photodisintegration of the deuteron the
differential cross
 section~(\ref{CMC}) is viewed as a function of the lab
photon energy $E_\gamma$ and
 angle $\Theta_p$ between incoming-photon and
outgoing-proton 3-momenta in the
 c. m. system of the np pair. One can obtain
the following kinematic relations
 between the laboratory photon energy and
variables in the c.~m.~frame:

\begin{eqnarray}
|{\bf p}|=\sqrt{\frac{s}{4}-m^2},\quad
s=M_d^2+2E_\gamma M_d,\quad \omega=\frac{M_d}{\sqrt{s}}E_\gamma.
\label{kin-s}
\end{eqnarray}

Following the Ref.~\cite{KorSheb} the invariant amplitude ${\cal M}_{fi}^\mu$
can be written in terms of the BS amplitude of the initial~(\ref{BS-disc})
and final~(\ref{BS-cont}) states as follows:
\begin{equation}
{\cal M}_{\text{fi}}^\mu=\frac{1}{4\pi^3}\int\!\! d^4k d^4l
\bar{\chi}_{Sm_s}(l;pP) \Lambda^\mu(l,k;P,K)\psi_{M}(k;K),
\label{matrix-element}
\end{equation}

\noindent where $S=0,1$ is the total spin of the np pair and $m_s$ is its
projection on to the $Z$ axis, $M$ is a projection the total angular momentum
of the deuteron; $\Lambda_\mu$ denotes the Mandelstam vertex which determines
the EM interaction with 2N system in the framework of the BS formalism.

Let us make some remarks on current conservation. The Mandelstam vertex
consists of one- and two-body parts, $\Lambda_\mu(p,k;P,K)=
\Lambda_\mu^{[1]}(p,k;P,K)+\Lambda_\mu^{[2]}(p,k;P,K)$. The second part of
the Mandelstam vertex  determines two-body contributions of the conserved
EM current. The specific form of $\Lambda_\mu^{[2]}(p,k;P,K)$ depends on a
given model for the interaction kernel in the BS equation and it cannot be
associated with the pair and meson exchange currents in the NR approach.
The gauge independence of EM current transition matrix element,
$q\cdot{\cal M}_{fi}=0$, will be fulfilled if the Mandelstam current meets
the following relations:
\begin{eqnarray}
&&iq\cdot\Lambda^{[1]}(p,k;P,K)=\\
&&\left\{
          \pi_p(1)\delta\left(p-k-\frac{q}{2}\right)
    \left[S^{(1)}\left(\frac{K}{2}+k\right)^{-1}
         -S^{(1)}\left(\frac{P}{2}+p\right)^{-1}
   \right]S^{(2)}\left(\frac{K}{2}-k\right)^{-1}\right.\nonumber\\
&&+
  \left.
         \pi_p(2)\delta\left(p-k+\frac{q}{2}\right)
   \left[S^{(2)}\left(\frac{K}{2}-k\right)^{-1}
        -S^{(2)}\left(\frac{P}{2}-p\right)^{-1}
  \right]S^{(1)}\left(\frac{K}{2}+k\right)^{-1}\right\}
\label{conserv}\end{eqnarray}

\noindent and
\begin{eqnarray}
&&iq\cdot\Lambda^{[2]}(p,k;P,K)=\nonumber\\
&&\sum\limits_{l=1,2}
\left[\pi_p(l){\cal V}\left(p+(-1)^l\frac{q}{2},k;K)\right)
             -{\cal V}\left(p,k-(-1)^l\frac{q}{2};K)\pi_p(l)\right)\right],
\label{conserv-1}\end{eqnarray}

\noindent where $S^{(l)}(p)$ is the fermion propagator and
$\pi_p(l)=1/2[1+\tau_z(l)]$ is the projector on to proton state.  Moreover
the BS amplitudes for the initial and final states have to satisfy to the BS
equations with the same interaction kernel.

In Ref.~\cite{RT5} it was shown that the gauge independence
condition for the elastic electron-deuteron scattering amplitude
is fulfilled in the impulse approximation  for many-rank separable BS kernels,
implying that $q\cdot{\cal M}_{fi}^{[2]}=0$. In case of the
deuteron breakup in the final 2N state isospin $I=1$ states are
present, yielding a nonvanishing isovector contribution,
$q\cdot{\cal M}_{fi}^{[2]}\neq0$. Consequently the two-body
Mandelstam current operator should be added to guarantee the gauge
independence.

\section{Plane wave one-body approximation}
\label{sec:pwia}

The amplitude~(\ref{matrix-element}) contains the FSI
contributions of the final np pair. This means that the half off
mass shell NN scattering $T$-matrix is needed,
v.~s.~Eqn.~(\ref{BS-cont}). In this paper rescattering
contributions and the pair processes in the framework of the BS
formalism will be not taken into account. The neglect of FSI is a
shortcoming of our present work. At present this is bound up with
computational difficulties. In a forthcoming paper will be
considered the FSI interactions from $J=0$, $^1S_0$, and $J=1$,
$^3S_1$--$^3D_1$, channels and work on the pair processes is in
progress. Full and detailed analysis of the FSI cannot clearly be
avoided without reconsidering an entirely different interaction
kernel.

In this investigation we confine ourselves to the simplest
 contribution to
the EM current, where the photon couples to one of
 the two nucleons in the
deuteron and the FSI between the
 outgoing nucleons are dropped --- the
plane-wave one-body approximation.

\subsection{The BS amplitude for the continuous spectrum}

According to the approximations we made the BS amplitude of the final state
given in Eqn.~(\ref{BS-cont}) is the antisymmetric combination of two free
Dirac positive energy spinors \footnote{
We use the covariant normalization of the Dirac spinors, $u^+ u= 2E$}:
\begin{eqnarray}
&&\chi_{Sm_s}(k_0,{\bf k};\sqrt{s}{\bf p})=4\pi^3\delta(k_0)
\label{free-BS}\\
&&\times\left[
\chi_{Sm_s}({\bf p})
(\eta_0+\eta_1)\delta^{(3)}({\bf k}-{\bf p})+
(-1)^{S+1}
\chi_{Sm_s}(-{\bf p})
(\eta_0-\eta_1)\delta^{(3)}({\bf k}+{\bf p})\right],\nonumber
\end{eqnarray}

\noindent where $\chi_{Sm_s}({\bf p})=
\sum\limits_{\lambda_p,\lambda_n=\pm\case{1}{2}} C_{\frac12 \lambda_p \frac12
\lambda_n}^{Sm_s} u_{\lambda_p}({\bf p}) u_{\lambda_n}(-{\bf p})$; $\eta_0$
and $\eta_1$ stands for isopin singlet and triplet functions respectively.
Since the outgoing nucleons are on mass shell, we have constraint $P\cdot
\hat p=0$ which keeps the relative energy $p_0$ to be equal to zero in the
rest frame of the np pair.

\subsection{The BS amplitude for the bound state}

The BS equation for the deuteron is solved in its rest frame.
Since the vertex function in Eqn.~(\ref{matrix-element}) is
referred to a moving frame, it has to be boosted to its rest
frame. The Lorentz transformation between the laboratory and
c.~m.~frames is given by $K^{\mu}={\cal L}^\mu_\nu K^{\nu}_{(0)}$,
$P^{\mu}={\cal L}^\mu_\nu P^{\nu}_{(0)}$, where $K^{\nu}_{(0)}$ and
$P^{\nu}_{(0)}$ --- 4-momenta of the deuteron and np-pair in their
rest frame, respectively. As only boost along the $Z$ axis is
needed, an explicit expression for the matrix ${\cal L}^\nu_\mu$
is given by
\begin{equation}
{\cal L}^\nu_\mu=\left(
\begin{array}{cccc}
\sqrt{1+\eta} & 0 & 0 & -\sqrt{\eta} \\
0 & 1 & 0 & 0 \\
0 & 0 & 1 & 0 \\
-\sqrt{\eta} & 0 & 0 & \sqrt{1+\eta}
\end{array}
\right),
\label{boost}\end{equation}

\noindent where the Lorentz-factor $\gamma\equiv\sqrt{1+\eta}$ and
$v\gamma\equiv\sqrt{\eta}$, with $v$ being the velocity of c.~m. frame in the
laboratory frame are defined in terms of the dimensionless boost
parameter $\eta$:
\begin{eqnarray}
\sqrt{\eta}=\frac{E_\gamma}{\sqrt{s}},\quad
\sqrt{1+\eta}=\frac{E_\gamma+M_d}{\sqrt{s}}.
\nonumber\end{eqnarray}

\noindent Under the boost the vertex function is transformed according to the
general rule of the transformation of spinor amplitudes
\begin{equation}
\Gamma_M(k;K)=\Lambda({\cal L})\Gamma_M({\cal L}^{-1}k;K_{(0)})
\label{boost-vertex}
\end{equation}

\noindent with $\Lambda({\cal L})= \Lambda^{(1)}({\cal L})\Lambda^{(2)}({\cal
L})$, where $\Lambda^{(l)}({\cal L})$ is the boost operator in the spinor
space of $l$th nucleon corresponding to ${\cal L}$:
\begin{eqnarray}
\Lambda^{(l)}({\cal L}) =
\left(\frac{1+\sqrt{1 + \eta}}{2}\right)^{\frac{1}{2}} \left(1 +
\frac{\gamma_{0}\gamma_{3}\sqrt{\eta}}{1 + \sqrt{1 +
\eta}}\right)^{(l)}.
\label{boost1}
\end{eqnarray}

\noindent At $\eta\to 0$ the matrix~(\ref{boost}) and operator~(\ref{boost1})
turn to unity, i.~e. ${\cal L}\to I$ and $\Lambda\to I$. This corresponds to
the static limit for the BS amplitude and takes place on threshold of
deuteron photodisintegration, a case of $\frac{E_\gamma}{m}\ll 1$ (v.~s.
Eqn.~(\ref{kin-s})). In the Eq.~(\ref{boost1}) the $\gamma_0\gamma_3$-term
affects the spin degrees of freedom of the BS amplitude.

\subsection{The electromagnetic vertex}

One-body part of the Mandelstam vertex has the form
\begin{eqnarray}
\Lambda_\mu^{[1]}(p,k;P,K)&=&i\delta^{(4)}
\left(p-k-\frac{q}{2}\right)
\Gamma^{(1)}_\mu\left(\frac{P}{2}+p,\frac{K}{2}+k\right)
S^{(2)}\left(\frac{K}{2}-k\right)^{-1}\nonumber \\
&+&i\delta^{(4)}\left(p-k+\frac{q}{2}\right)
\Gamma^{(2)}_\mu\left(\frac{P}{2}-p,\frac{K}{2}-k\right)
S^{(1)}\left(\frac{K}{2}+k\right)^{-1},\label{tn}
\end{eqnarray}

\noindent where $\Gamma_\mu^{(l)}(p,k)$ is off-mass-shell $\gamma$NN vertex
for the $l$th nucleon. This is common problem of direct application of the BS
equation to the NN interaction. The consequence of gauge constraints for
off-shellness in the EM vertices have been recently considered in
Ref.~\cite{nagor-off} (see also references therein).  In our case we deal
with half off-mass shell EM vertex. This type of vertex occurs in the
$(e,e^\prime N)$ reaction, e.~g. nucleon knockout and inclusive electron
scattering, when the initial nucleon is taken to be bound (off-mass-shell)
and the knocked-out nucleon is assumed to be in physical state.  It is shown
in Ref.~\cite{nagor-off} that the off-shell behavior of the EM vertex of the
nucleon in a real Compton scattering on a free nucleon does not play a role
as consequence of gauge invariance.

Thus we deal with well-known on-mass-shell version of the EM vertex:
\begin{eqnarray}
&&\Gamma_\mu^{(l)}(q) = \gamma_{\mu}\left(F^{(s)}_{1}(q) + \tau^{(l)}_{3}F^{(v)}(q)\right)
+ \frac{i}{2m}\sigma_{\mu\nu}q^{\nu}\left(F^{(s)}_{2}(q) + \tau^{(l)}_{3}
F^{(v)}_{2}(q)\right),\label{BN}
\end{eqnarray}
\noindent where $F^{(s,v)}_{1,2}(q)$ is the isoscalar  and
isovector Pauli-Dirac form factors of the $l$th nucleon,
normalized as: $F^{(s)}_{1}(0)=\case{1}{2}$,
$F^{(s)}_{2}(0)=\case{\kappa_p+\kappa_n}{2}$,
$F^{(v)}_{1}(0)=\case{1}{2}$,
$F^{(v)}_{2}(0)=\case{\kappa_p-\kappa_n}{2}$ with the anomalous
part of the proton and neutron magnetic momenta $\kappa_{p,n}$
respectively.

\begin{figure}[h]
\begin{center}
\epsfxsize=11cm \epsfysize=11cm \epsfbox{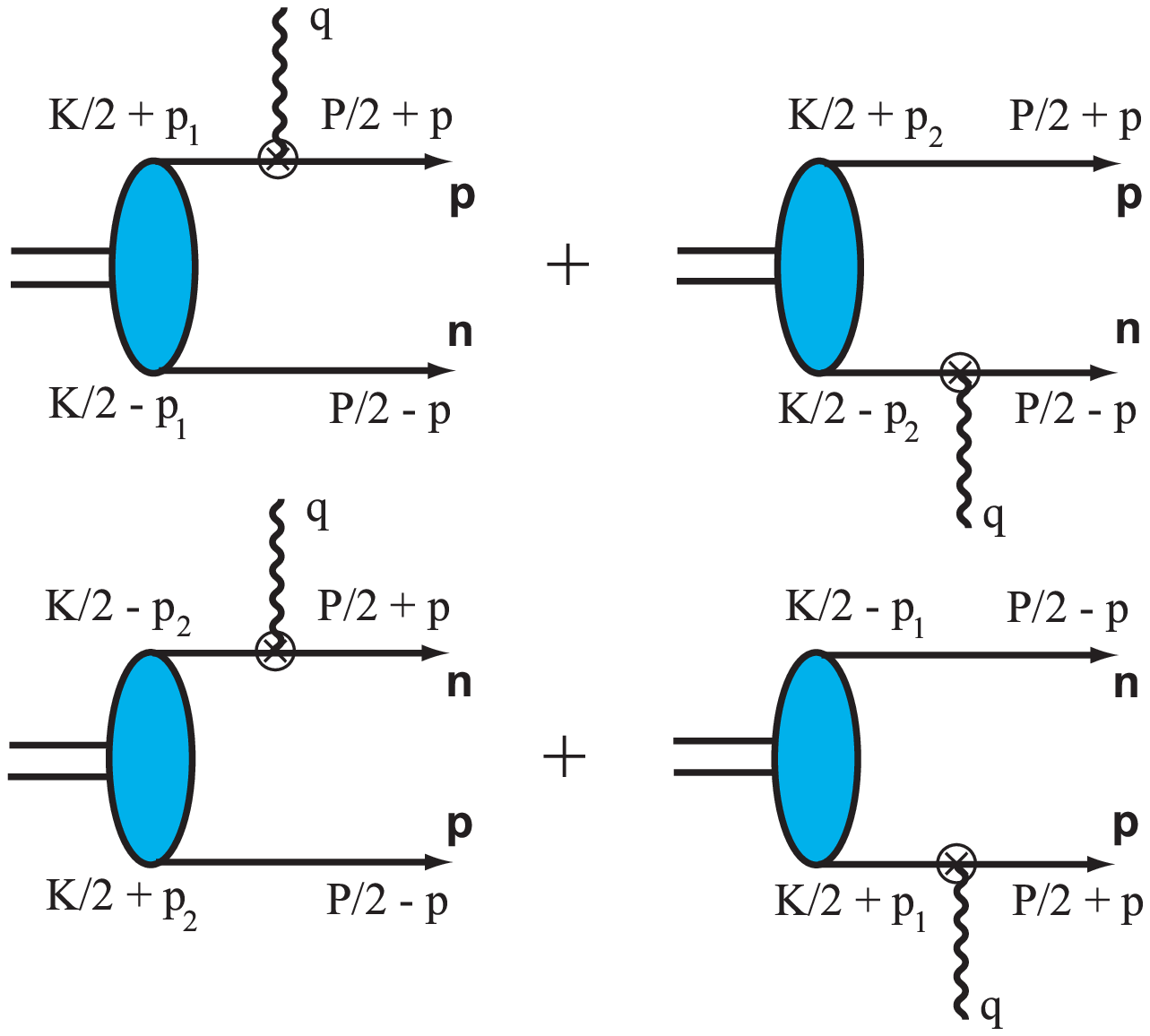} \caption{Kazakov
K.Yu}\label{fig1}
\end{center}

\end{figure}

Employing the written transformation laws, substituting
Eqns.~(\ref{free-BS}), (\ref{boost-vertex}), (\ref{tn}) to
Eqn.~(\ref{matrix-element}) and integrating over intermediate 4-momentum, we
arrive at an expression in terms of the c.~m. vertex function and propagators
\begin{equation}
{\cal M}^{\mu}_{\text{fi}}
=\sum_{l=1,2}
\bar \chi_{Sm_s}^{(0)}(0,{\bf p};\sqrt{s}{\bf p})
\Gamma^{(l)}_\mu(q^2=0)\Lambda({\cal L})
S^{(l)}(k;K_0)\Gamma_M(k_{0l},{\bf k}_l;K_0),
\label{AMPL}
\end{equation}

\noindent where $k_l={\cal L}^{-1}(p+(-1)^l\frac{q}{2})$. All
possible contributions to the transition matrix element in the
PWOA are depicted in Fig.~\ref{fig1}. Since both the initial and
final 2N states are antisymmetric the four diagrams are identical.

In the PWOA it is seen that the matrix element is proportional
directly to the deuteron vertex function taken at specific value
of energy-momentum. As the relative 4-momentum $k_l$ is restricted
by energy-momentum conservation in a photon-nucleon vertex, the
relative energy $k_{0l}$ and 3-momentum ${\bf k}_l$ variables
depend on the photon energy in the lab. frame and the
dimensionless boost parameter $\eta$:
\begin{eqnarray}
&&k_{0l}=\sqrt{\eta}\,|{\bf p}_{||}| +(-1)^l\frac{E_\gamma}{2},
\nonumber\\
&&{\bf k}_{\perp l}={\bf p}_\perp,\quad \omega=M_d\sqrt{\eta},\label{k01}\\
&&{\bf k}_{|| l}=\sqrt{1+\eta}\,{\bf p}_{||}
+(-1)^l\frac{\bf q}{2},\nonumber
\nonumber\end{eqnarray}

\noindent where indices $||$ and $\perp$ denotes the longitudinal and
transverse components of vector ${\bf k}$ with respect to direction of the
incoming photon 3-momentum ${\bf q}$, here $|{\bf q}|=E_\gamma$. The
situation is illustrated on Fig.~2. For photon energies $E_\gamma\leq
0.2~\text{GeV}$ one is probing the energy-momentum distribution of the bound
nucleons in the deuteron where the high-momentum `tails' of the nucleonic
states in deuteron  are especially relevant. For given c.~m. angles
$\Theta_p$ it is found that both the relative energy, which accounts for the
retardation in the vertex function of the deuteron, and the modulus of the
3-momentum of the proton (neutron) rise strongly with $E_\gamma$. They are
smaller at forward scattering angles, and for other angles covering wide
bands from 100~MeV to 2~GeV and from 500~MeV to 2.5~GeV.

Introducing the deuteron state components, the 2N continuum amplitude  in the
 matrix representation (for details we refer to the paper~\cite{KapKaz}), the
transition matrix element can be evaluated calculating traces of
$\gamma$-matrix expressions:

\begin{eqnarray}
{\cal M}_{\text{fi}}^\mu=
&-&Sp\left(\bar\chi_{Sm_s}({\bf p})
\Gamma_{p,\mu}\Lambda({\cal L})
S(\frac{K_0}{2}+k_1;K_0)
\Gamma_{M}(k_{1};K_0)\Lambda({\cal L}^{-1})\right)\nonumber\\
&-&
Sp\left(\bar\chi_{Sm_s}({\bf p})\Lambda({\cal L})\Gamma_{M}(k_{2};K_0)
\tilde{S}(\frac{K_0}{2}-k_2;K_0)\Lambda({\cal L}^{-1})
\Gamma_{n,\mu}\right)
,\label{TTY}
\end{eqnarray}

\noindent  with
\begin{eqnarray}
&&\bar\chi_{1m_s}({\bf p})=
\frac{{\cal N}_{\bf p}^2}{2\sqrt{2}}
(m-\widehat{p}_{2})\widehat{\xi}^{\ast}_{m_{s}}(1+\gamma_{0})(m+
\widehat{p}_{1}), \\ \nonumber
&&\bar\chi_{00}({\bf p})=\frac{{\cal N}_{\bf p}^2}{2\sqrt{2}}
(m-\widehat{p}_{2})\gamma_{5}(1 + \gamma_{0})(m +
\widehat{p}_{1}),
\end{eqnarray}

\noindent where $\xi_{m_s}$ is a polarization 4-vector with
the following  completeness and orthogonality relations
\begin{eqnarray}
\sum\limits_{m_s=-1}^{+1}\xi_{m_s}^\mu\xi_{m_s}^{\nu *}
=-g^{\mu\nu}+\frac{P^\mu P^\nu}{s},\qquad \xi\cdot P=0,
\label{sum-xi}\end{eqnarray}

\noindent and normalization constants ${\cal N}_{\bf p}$ and
vectors $p_{1,2}$ are defined in Eqn.~(\ref{3s1}).

Since the EM nucleon form factors can be taken by their on-shell form,
 we have   for the charge-current operator
\begin{eqnarray}
\Gamma_{p,\mu}=\gamma_\mu+\frac{i\kappa_{p}}{2m}\sigma_{\mu\nu}q^{\nu},\qquad
\Gamma_{n,\mu}=\frac{i\kappa_{n}}{2m}\sigma_{\mu\nu}q^{\nu}.
\end{eqnarray}

\noindent The fermion propagator in Eqn.~(\ref{TTY})
\begin{eqnarray}
\tilde{S}(k) = \frac{\widehat{k}-m}{k^{2}-m^{2}+i\epsilon}
\label{GP}
\end{eqnarray}

\noindent is connected with the propagator $\tilde{S}$ by $\tilde{S}=-C S^T
C$, where $C=i\gamma^2\gamma^0$.

Using the expression~(\ref{TTY}), we find that the differential
photo-absorption cross section can be written as
\begin{equation}
\label{final-cross-section}
 \frac{d\sigma}{d\Omega_p}=\frac{d\sigma_0}{d\Omega_p}+\frac{d\sigma_{\text{SP}}}{d\Omega_p},
\end{equation}

\noindent where the $\sigma_0$ is the part of the cross section
which makes up the shape of the angular distributions
\begin{equation}\label{c-s-0}
\frac{d\sigma_0}{d\Omega_p}=\frac{\alpha}{4\pi
s}\left(\frac{1+\sqrt{1+\eta}}{2}\right)^{2}\sum\limits_{S=0,1}\mid
{X}_0^S\mid^{2}
\end{equation}

\noindent and the $d\sigma_{\text{SP}}$ accounts for the effect of
the boost on the spin degrees of freedom (the spin precession) of
the nucleons
\begin{equation}\label{spin-boost}
\frac{d\sigma_{\text{SP}}}{d\Omega_p}=\frac{\alpha}{4\pi
s}\left(\frac{1+\sqrt{1+\eta}}{2}\right)^{2}\sum\limits_{S=0,1}
\mid {\cal M}^S_{\text{SP}}\mid^{2}
\end{equation}

\noindent with the square modulus of the amplitude ${\cal
M}^S_{\text{SP}}$ is given by
\begin{eqnarray}
{\mid{\cal M}_{\text{SP}}^S\mid}^{2}&=&2\beta Re(X_0^S
X^{S\ast}_{1})\nonumber\\ &+&\beta^{2}\left(\mid X_{1}^S\mid^{2}
-2Re(X_{0}^SX^{S\ast}_{2})\right)-2\beta^{3}Re(X_{1}^SX^{S\ast}_{2})
+\beta^{4}\mid X_{2}^S\mid^{2}\label{final-mat}
\end{eqnarray}

\noindent with $\beta=\case{\sqrt{\eta}}{1+\sqrt{1+\eta}}$ and the
amplitudes $X_i^S$ $(i=0,1,2)$ are expressed as
\begin{eqnarray}
X_0^S&=&Sp\biggl(\bar\chi_{Sm_s}({\bf p})
\,\Gamma_{p,\,\lambda}\,S^{(1)}(s_{1};K_0)\Gamma_{M}(k_{1};K_0)\biggr)
+~p\leftrightarrow~n,\nonumber \\
X_1^S&=&Sp\biggl(\bar\chi_{Sm_s}({\bf p})
\,\Gamma_{p,\,\lambda}\,\gamma_{0}\widehat{n}_{3}
S^{(1)}(s_{1};K_0)\Gamma_{M}(k_{1};K_0)\biggr)\nonumber\\
&-&Sp\biggl(\bar\chi_{Sm_s}({\bf p})
\,\Gamma_{p,\,\lambda}\,S^{(1)}(s_{1};K_0) \Gamma_{M}(k_{1};K_0)
\gamma_{0}\widehat{n}_{3}\biggr)+~p\leftrightarrow~n,\label{XXX}\\
X_2^S&=&Sp\biggl(\bar\chi_{Sm_s}({\bf p})
\,\Gamma_{p,\,\lambda}\,\gamma_{0}\widehat{n}_{3}
S^{(1)}(s_{1};K_0)\Gamma_{M}(k_{1};K_0)\gamma_{0}\widehat{n}_{3}\biggr)
+~p\leftrightarrow~n\nonumber
\end{eqnarray}

\noindent with $n_{3} = (0,0,0,1)$ denoting a unit vector and
$s_1=K_0/2-k_1$.

These are the general expressions for the deuteron
photodisintegration cross section~(\ref{CMC}) in the PWOA.  Since
we omit the two-body contribution to the transition matrix
element, we does not preserve the gauge independence of the
amplitude. Thus averaging over photon polarizations, we make use
of Coulomb gauge, $\varepsilon^0=0$,
$\mbox{{\boldmath$\varepsilon$}}\cdot {\bf q}=0$ with the
completeness relation of the form
\begin{eqnarray}
\sum\limits_{\lambda=\pm 1}
(\varepsilon_\lambda)_i^*(\varepsilon_\lambda)_j
=\delta_{ij}-\frac{q_iq_j}{{\bf q}^2},\qquad i,j=x,y.
\label{sum-epsilon}\end{eqnarray}

All $\gamma$-matrix expressions in the matrix elements in Eq.~(\ref{XXX}) and
the square modulus of the amplitude in Eq.~(\ref{final-mat}) are evaluated
with the computer algebraic program REDUCE. When summing over nuclear
polarizations, we make use of the relations~(\ref{sum-e}) and~(\ref{sum-xi}).

\section{Analysis of relativistic effects}
\label{sec:analysis}

Now we are in a position to do the final computations. The results for the
angular distributions for deuteron photodisintegration at four different
photon energies $E_\gamma$ in the laboratory frame are depicted in
Fig.~\ref{fig3}. The cross section is calculated in the framework of the BS
equation with the separable interaction kernel (with two different strength
of $D$-state, $P_2$ = 4~\% and $P_2$ = 6~\%). This interaction kernel is
similar to that employed in the calculations of the deuteron EM elastic form
factors~\cite{RT5}.  For simplicity, though it might be important, in our
calculations we disregard negative-energy partial states of the deuteron
vertex function. These will be considered in detail in the BS formalism for a
meson exchange interaction kernel.

In Fig.~\ref{fig3} it is seen that the resultant curves reproduce shapes
of angular distributions~\cite{ArSan}.  Above the threshold it is an
almost perfect $\sin^2\Theta_p$ behavior, which corresponds to $E1$
transition to the $^3P_0$ np continuum state. At higher photon energies
the overall magnitude of the cross section rapidly falls off and the
maximum is shifted from 90$^\circ$ to 70$^\circ$ and, further, to
60$^\circ$. The whole distributions, which are dominated by magnetic
transitions, become flatten, the ratio of the forward cross section to the
maximum decreases. The role of the $D$-state becomes more pronounced. But
notwithstanding these similarities, the theory appears systematically less
than the experimental distributions.  At $E_\gamma$=20~MeV it is less by
factor 1.4, at $E_\gamma$=100~MeV by factor 3, in the $\Delta$-resonance
region it is expectedly lower by almost factor 10, and at
$E_\gamma$=500~MeV by factor 3 with respect to $D$-state weight $P_2$ =
6~\%. In the NR approach, incorporating meson-nucleon degrees of freedom,
a good deal of discrepancy between the experiment and theory is diminished
by contributions from meson-exchange currents and isobar configurations.

\subsection{Approximate calculations}

Despite the fact that our theoretical results does not describe the
experimental data, we are able to make definitive statements concerning the
relative importance of the various relativistic effects. We distinguish the
following classes of the relativistic effects in our consistent treatment:
relativistic kinematics and EM current operator; relativistic NN dynamics,
which forbids instantaneous interactions, and leads to the relative energy
dependence of the deuteron vertex function (retardation), the boost
transformation affecting the internal variables of the BS amplitude of the
deuteron (Lorentz contraction), the part of the cross section denoted as
$\frac{d\sigma_0}{d\Omega}$, and its spin degrees of freedom (spin
precession), the cross section $\frac{d\sigma_{\text{SP}}}{d\Omega}$.

\begin{figure}[t]
\vspace{-5cm}
\begin{center}
\vbox to 22cm { \epsfxsize=12cm \epsfysize=22cm
\hspace*{1cm}\epsfbox{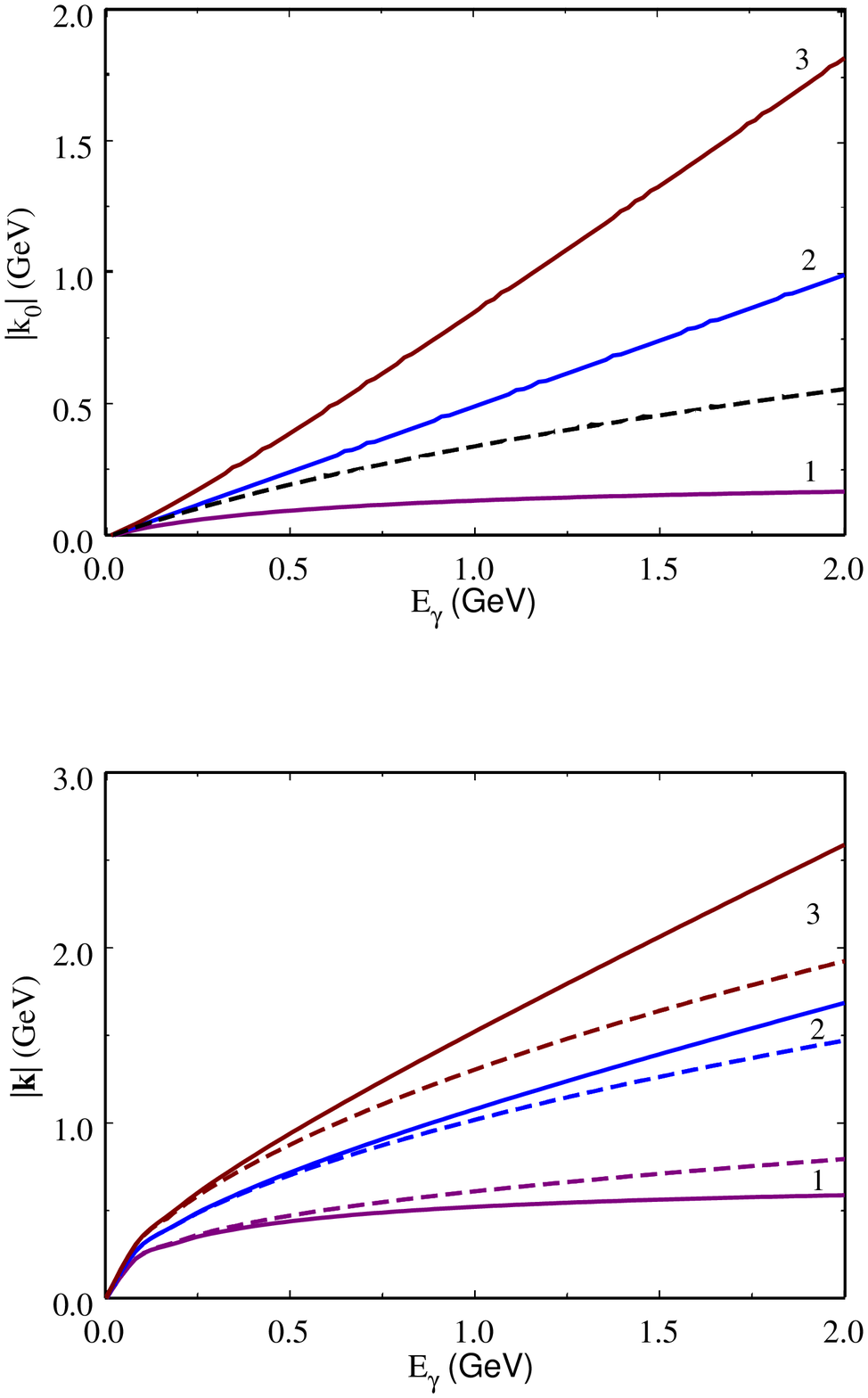}}
\caption{Kazakov K.Yu.} \label{fig2}
\end{center}
\end{figure}

\begin{figure}[t]
\vspace*{-0cm} \vbox to 20cm{\epsfxsize=19cm
\epsfysize=20cm \epsfbox{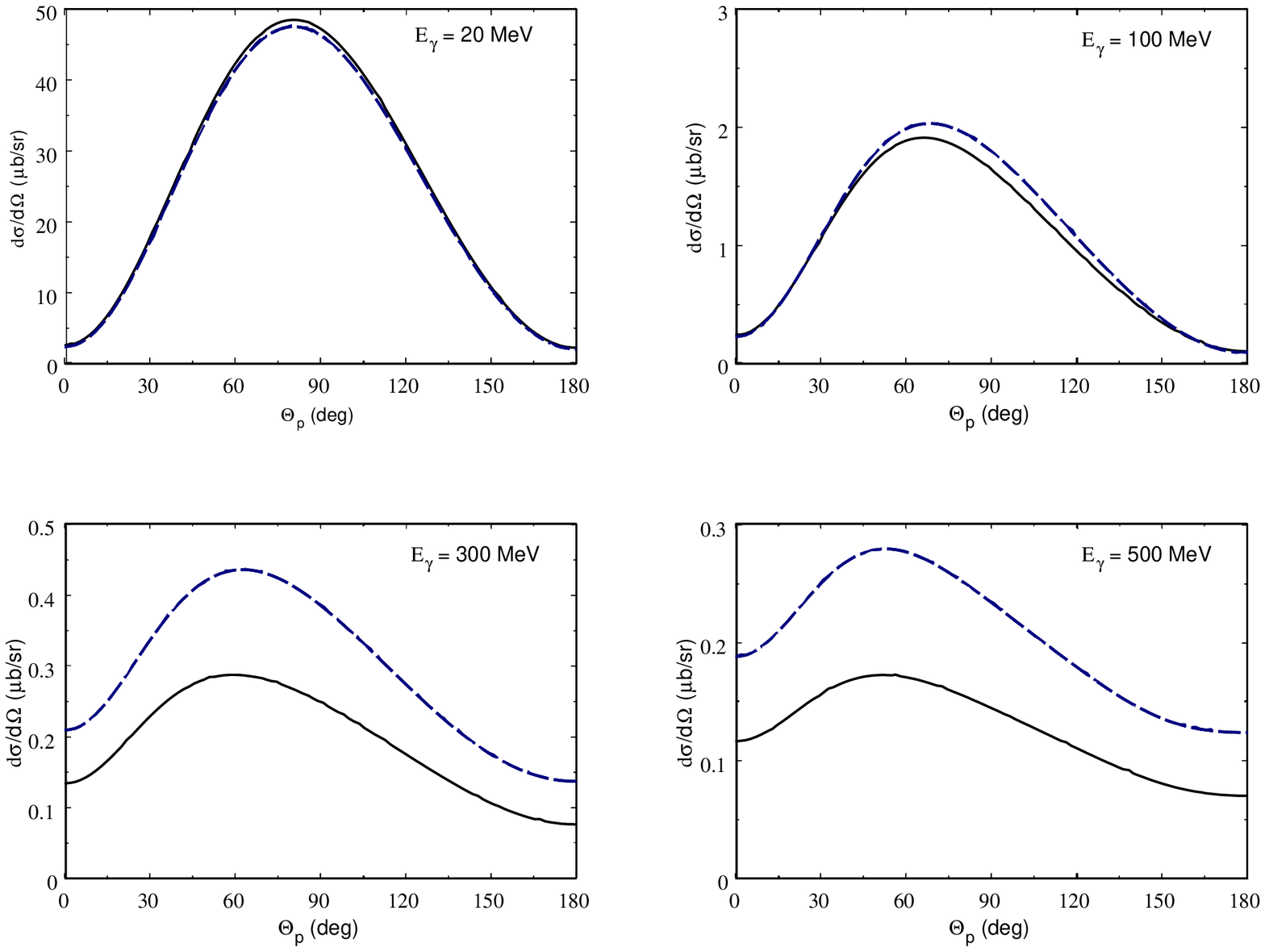}} \caption{Kazakov
K.Yu.}\label{fig3}
\end{figure}

We start discussion with evaluation of the size of contributions due to
retardation, Lorentz contraction and spin precession. Let us consider a
number of approximate calculations with respect to the exact positive-energy
BS calculations:
\begin{enumerate}
\item First of all we perform the static approximation (BS-SA) to the BS
cross sections, which amounts to neglecting the boost on the
arguments of the deuteron vertex function and one-particle
propagator, see Eq.~(\ref{AMPL}). It is achieved by putting the
boost parameter $\eta=0$ in the deuteron vertex function
$\left.\Gamma_L(k_{0l},{\bf k}_l)\right|_{\eta=0}=
\Gamma_L(p_{0l},{\bf p}_l)$, where
$p_{0l}=(-)^{l}\case{E_\gamma}{2}$ and ${\bf p}_l={\bf
p}+(-)^{l}\case{{\bf q}}{2}$ $(l=1,2)$. Kinematically this effect is
shown in Fig.~\ref{fig2}. The booster in
Eq.~(\ref{boost1}) is approximated by $\left.\Lambda({\cal
L})\right|_{\eta=0}=I$. This approximation excludes contributions
due to the Lorentz contraction and spin precession.

\item Moreover we pay special attention to investigation of the
influence of the boost on the nucleon relative energy in one-particle
propagator. As it is shown in Ref.~\cite{RT4} that this is the most
important relativistic contributions to the deuteron EM form factors. Here
we consider the case (BS-BR), which is the same as BS-SA, but includes the
boost on one-particle propagator $S^{(l)}(k_{0l},|{\bf p}_l|)$  due to
recoil, $k_{0l}=\sqrt{\eta}|{\bf p}|+(-1)^l\case{E_\gamma}{2}$.

\item Finally, in order to find out the relativistic correction
associated with the relative energy dependence in the matrix
elements, which is brought by the BS vertex function. We consider
the zero-order approximation (BS-ZO) for the vertex function,
i.~e. computing the radial parts of the vertex function in the
BS-BR approximation with $k_{0l}$ equal to zero.

\end{enumerate}

\begin{figure}[t]
\vspace*{-0cm} \vbox to 20cm{\epsfxsize=19cm
\epsfysize=20cm \epsfbox{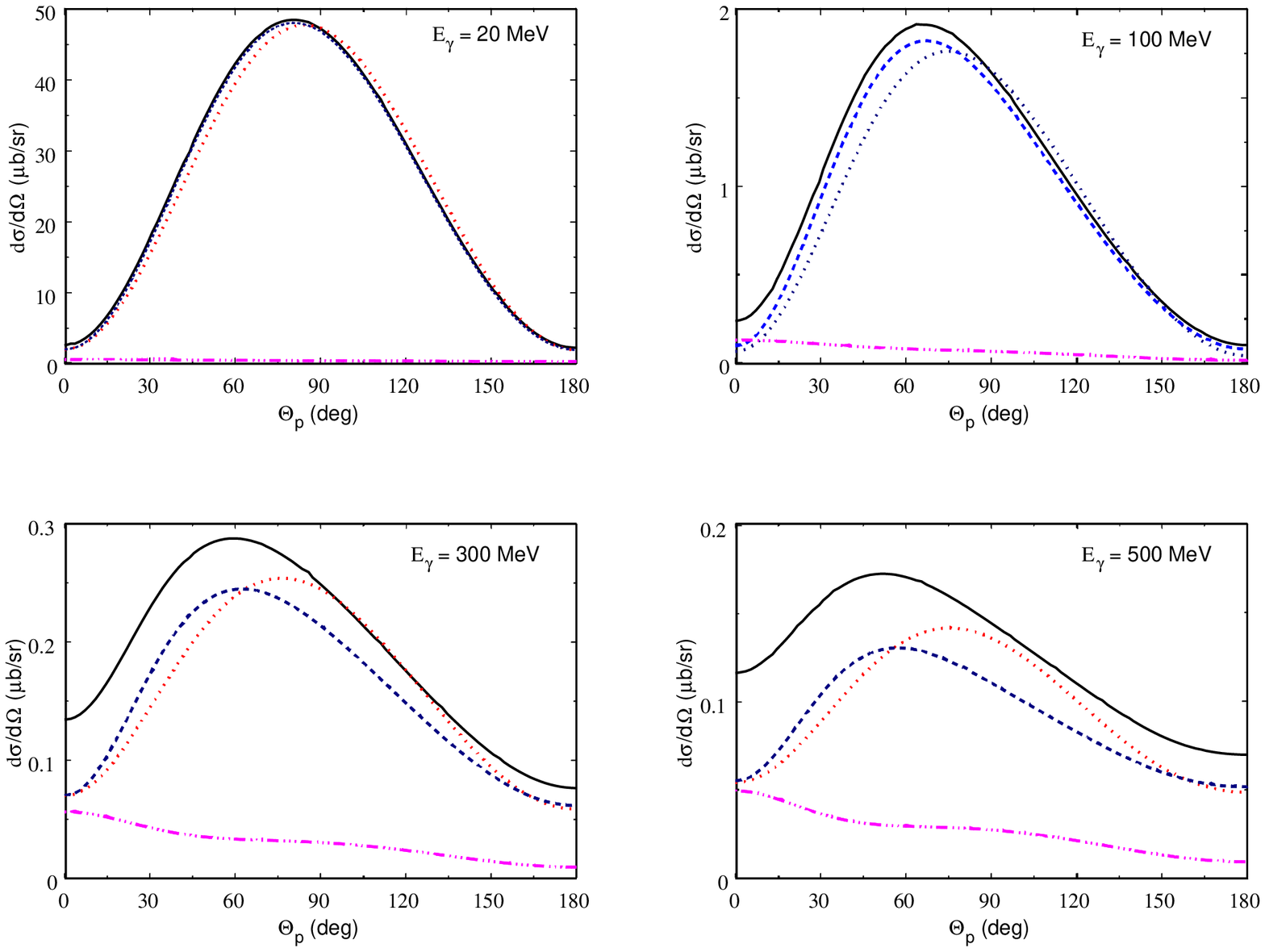}} \caption{Kazakov
K.Yu.}\label{fig4}
\end{figure}

In Fig.~\ref{fig4} we present the results for the angular distributions at
the same lab. photon energies as in Fig.~\ref{fig3}. The solid curve
corresponds to the exact calculation. We would like to stress the importance
of the effects concerning the booster. In Fig.~\ref{fig4} it is clearly seen
the role of the $d\sigma_{\text{SP}}$ cross section (3-dot-dash line) with
rise of the energy. It becomes pronounced at the scattering in the forward
semi-sphere, where it is almost the half of size of the cross section. One
should allow for such a contribution starting at the medium photon energies.
When added to the BS-BR approximation (dash line), the latter becomes to be
plausible for discussed $E_\gamma$ region. The BS-BR approximation, which
includes the boost on the one-particle propagator, gives the shape of the
angular distributions close to the exact ones. We can conclude that the BS-BR
approximation supplemented with $d\sigma_{\text{SP}}$ contribution accounts
for major relativistic effects in the cross section. On the other hand the SA
approximation (dot line) ceases to be reasonable for the $E_\gamma$ above
100~MeV, as it has a wrong position of cross section maxima.

\begin{figure}[t]
\vspace{-5cm}
\begin{center}
\vbox to 22cm { \epsfxsize=12cm \epsfysize=23cm
\hspace*{1cm}\epsfbox{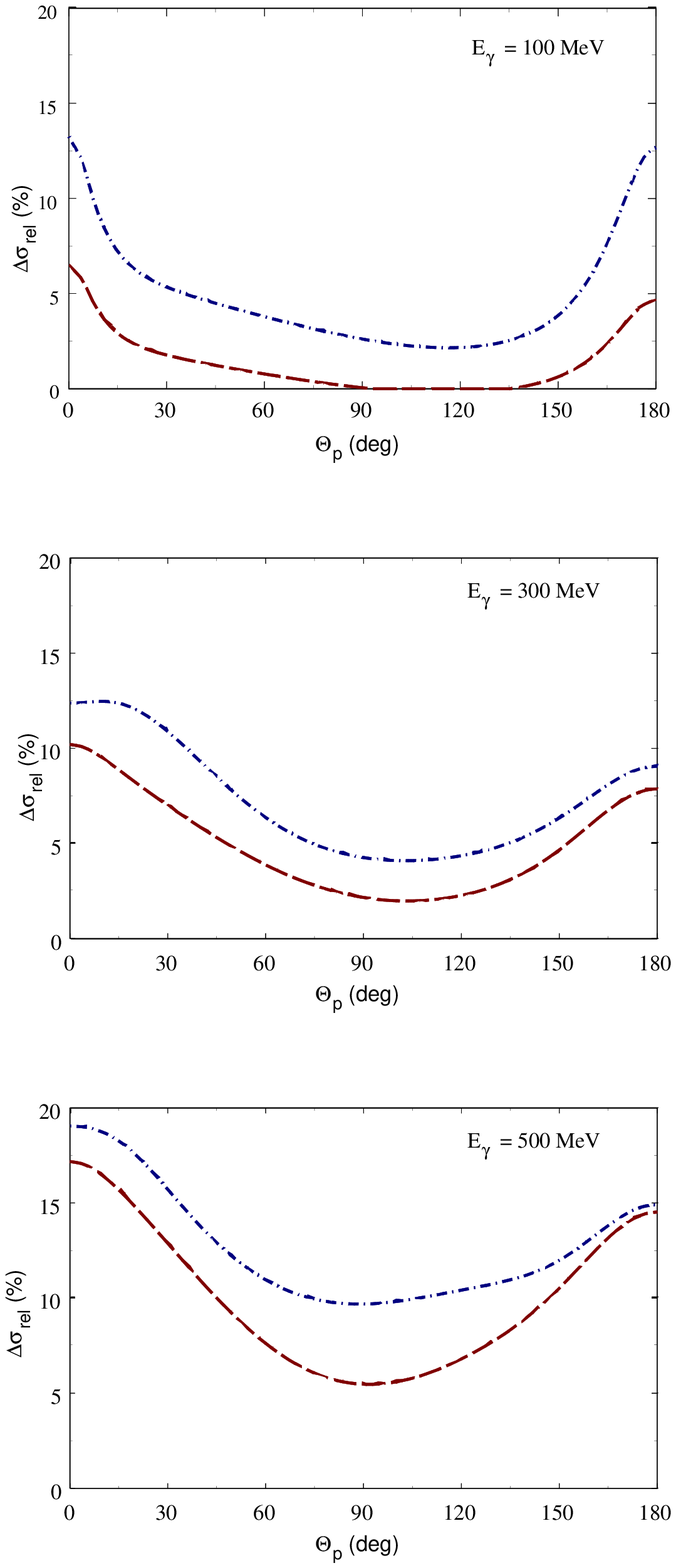}}
\caption{Kazakov K.Yu.} \label{fig5}
\end{center}
\end{figure}

We show the influence of the boost transformation in the arguments of the
initial-state vertex function. It is worthwhile to mention that the boost
leaves the arguments of the radial part of the vertex function unchanged (the
radial function $g_L(p;K)$ depends on the Lorentz invariants $K^2$, $p^2$ and
$p\cdot K$). It only has a direct bearing on its spin-orbital part. In
Fig.~\ref{fig5} it is displayed the relative deviation of the BS-BR (dash
line) and BS-ZO (dot-dash line) approximations from the BS cross section,
$d\sigma_0$. The deviation of the BS-BR cross section is due to the boost
effect on the orbital part of the deuteron vertex function. The respective
contribution is quite great especially at forward and backward proton c.~m.
angles. The effects of retardation is responsible for discrepancy between the
two curves in Fig.~\ref{fig5}. One can see that this is practically uniform
difference reaching up to 5~\%. If one ignores the dependence on the relative
energy in the deuteron vertex function, one comes up with the cross section
which is slightly smaller but keeps the same shape.

\subsection{Comparison with other approaches}

The BS formalism consistently accounts for the relativistic effects
associated with manifest Lorentz covariance of scattering amplitudes.  As far
as matter of relativity is concerned, we compare the exact results of the BS
framework with the following approaches.

\begin{enumerate}
 \item In the PWOA the ET approximation can be obtained immediately from the
 exact expressions replacing the BS deuteron bound state for ``$++$"-channels
 by the QP vertex function, which is solution of the 3-D QP equation with the
 BbS propagator~(\ref{propagator-BBS}) and with the refitted version of the
 separable interaction kernel Graz-II~\cite{RT5}. Essentially this
 approximation makes use of instantaneous interactions, i. e. with the zero
 relative time, or respectively, relative energy in the deuteron vertex
 function $g_L$ and the one-particle propagator.

\item A minimally relativistic approach employs the same QP vertex function
of the deuteron corresponding to solution of the bound state equation the
BbS propagator. This corresponds to the static limit for the ET
approximation (ET-SA). The approach incorporates the relativistic
kinematics and covariant form of the EM current matrix elements.

\item  A purely non-relativistic approach makes use wave function of the
deuteron, which is solution of the Schr\"dinger equation with NR Graz-II
separable potential.
\end{enumerate}

\begin{figure}[t]
\vspace{-5cm}
\begin{center}
\vbox to 22cm { \epsfxsize=12cm \epsfysize=23cm
\hspace*{1cm}\epsfbox{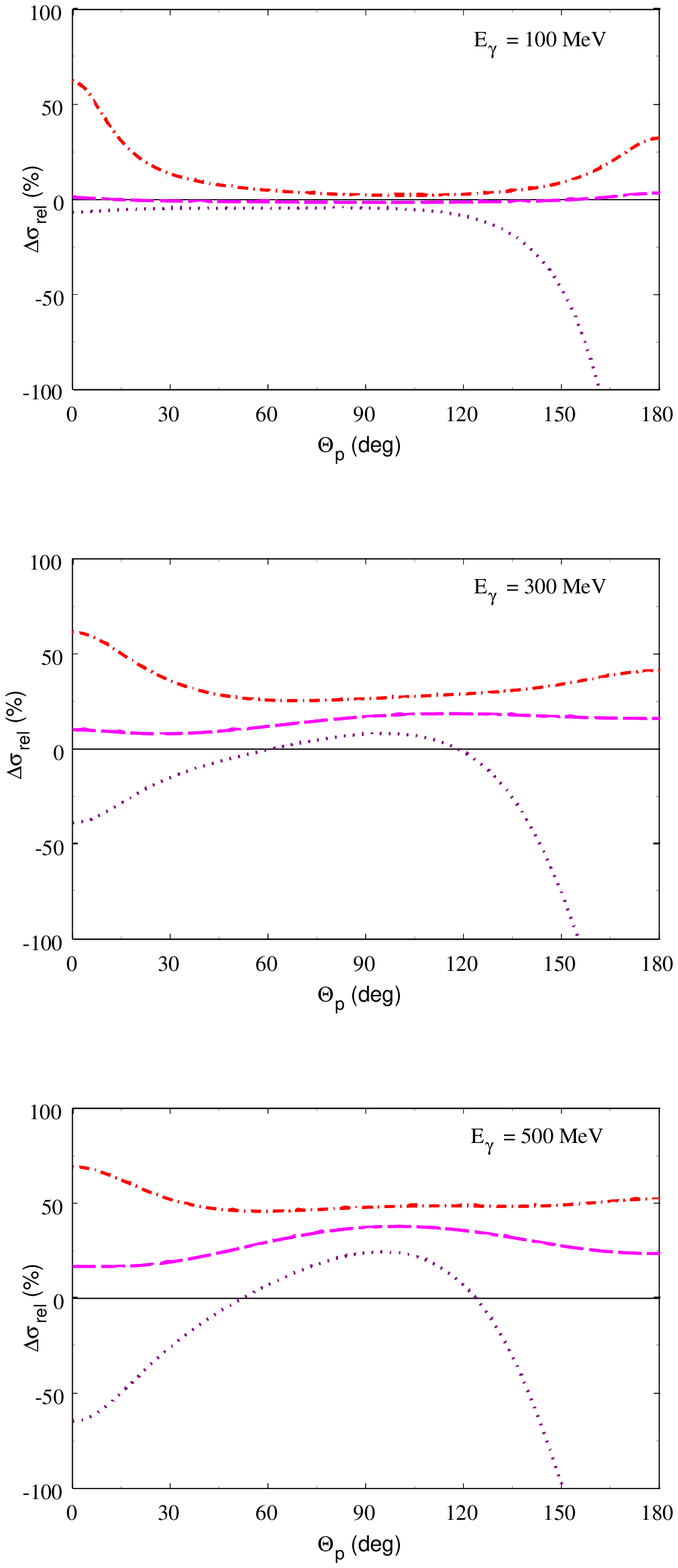}}
\caption{Kazakov K.Yu.} \label{fig6}
\end{center}
\end{figure}

In Fig.~\ref{fig6} we compare the above mentioned approaches with the
exact relativistic calculation. One can see that NR approach (dot line) is
very crude approximation for the photon energies greater that 100~MeV. The
minimally relativistic approach (dot-dash line) improves the situation.
By no means the NR approach is reasonable to describe deuteron
photodisintegration at high photon energies above the pion threshold. At
least one should include minimally relativistic corrections. The
conclusion is in accordance with the discussion in Ref.~\cite{ArSan},
where it was shown that should the relativistic effects not included, a
theory gave too much peaking of the differential cross section at
$\Theta_p$=0$^\circ$ and 180$^\circ$.

As one can see the overall sign of the relativistic contribution to the NR
cross section is negative, but it depends on given photon energy in the range
$\Theta_p$=60$\div$120$^\circ$.

Taking into account of Lorentz deformation (dash line in Fig.~\ref{fig6})
produces sizeable effect. It is seen that the ET approximation is a good
approximation at low photon energies. Almost equally at all proton
c.~m.~angles, the discrepancy between the exact and ET calculations reaches
about 10~\% at $E_\gamma$=300 MeV and 25 percent at the $E_\gamma$=500 MeV.

\subsection{Expansion of the relativistic model}

It is well-known that the usual way to include relativistic
effects is $\case{|{\bf p}|}{m}$- and $\case{\omega}{m}$-expansion
of the exact relativistic model, if one confines to the lowest
order correction $\case{|{\bf p}|^2}{m^2}$ and $\case{\omega}{m}$
beyond the NR amplitudes~\cite{ArSan}. This relativistic
correction is valid at the photon energies $E_\gamma$ up to few
hundred MeV and it seems to correspond to corrections of
spin-orbit type to the NR current operator.

We analytically performed approximation of the expression~(\ref{final-mat})
for the angular distributions in the limit $E_\gamma\ll m$. For the c.~m.
frame variables in the NR limit, corresponding to energies $E_\gamma\lesssim
100$~MeV, we have $|{\bf p}|\cong\sqrt{m(E_\gamma-\epsilon_d)}$, $\omega\cong
E_\gamma$ and the boost parameter $\sqrt{\eta}\cong 0$. Thus the matrix
element is expanded in powers of $\case{|{\bf p}|}{m}$ and $\case{\omega}{m}$
with $\case{|{\bf p}|}{m}\approx 0.3$ keeping the lowest order terms, recoil
effects and boost effects are neglected as well.  The result is that of the
non-covariant description of deuteron in the framework of the conventional NR
models incorporating relativistic effects in a $\case{|{\bf p}|}{m}$
expansion of a relativistic model~\cite{ArSan,chemtob}.

The differential cross section of a photon with energy $E_\gamma$ in the NR
framework is given by
\begin{eqnarray}
\frac{d\sigma_0}{d\Omega_p}=\frac{\alpha |{\bf p}|}{4\pi E_\gamma}
\sum\limits_{S=0,1}\overline{\mid{X_0^S}\mid^{2}},
\label{NR-cross}\end{eqnarray}

\noindent where the deuteron
break-up matrix elements are expressed in terms of the NR deuteron wave
function $\Psi_{M}({\bf k})$, the wave function of the 2N scattering states
$\Psi_{{\bf p}Sm_s}({\bf k})$ in the following form
\begin{eqnarray}
X_0^S=-\sum\limits_{l=1,2}
\int\frac{d{\bf k}}{(2\pi)^3}\bar\Psi_{{\bf p}Sm_s}({\bf k})
F^\lambda_l({\bf q})\Psi_{M}({\bf k}_l),\qquad (\lambda=\pm 1),
\label{NR-mat-el}\end{eqnarray}

\noindent where ${\bf k}_l={\bf p}-(-1)^l\case{{\bf q}}{2}$ and the EM
nucleon form factor is given by
\begin{eqnarray}
F^\lambda_l({\bf q})=(-1)^{(l+1)}
\frac{1+\tau_z^{(l)}}{2}
\frac{{\bf p}^\lambda}{m}+\frac{\kappa_s+\tau_z^{(l)}\kappa_v}{2}
\frac{i[\mbox{{\boldmath$\sigma$}}^{(l)}\times{\bf q}]^\lambda}{2m}
\nonumber\end{eqnarray}

\noindent with $\kappa_s=\case{1}{2}(\kappa_p+\kappa_n)$ and
$\kappa_v=\case{1}{2}(\kappa_p-\kappa_n)$.

In the PWOA one can reduce square of the
amplitude~(\ref{NR-mat-el}) in Eq.~(\ref{NR-cross}) to `textbook'
formulae which reproduces the shape of a angular distribution but
misses its absolute size:
\begin{eqnarray}
d\sigma_0&=&
\frac{{\bf p}^2}{2m^2}\sin^2\Theta_p
\Biglb(U_1^2+W_1^2\Bigrb)+\frac{E_\gamma^2}{4m^2}
\biggl\{\kappa_p^2\Biglb(U_1^2+W_1^2\Bigrb)+\kappa_n^2\Biglb(U_2^2+W_2^2\Bigrb)
\biggr.\nonumber\\
&+&\frac{\kappa_p\kappa_n}{3}\biggl[2U_1U_2
-\frac{U_1W_2+U_2W_1}{\sqrt{2}}\Biglb(1+3\cos(2\Theta_p)\Bigrb)\biggr.
+\biggl.\biggl.\frac{W_1W_2}{2}\Biglb(5+3\cos(2\Theta_p)\Bigrb)\biggr]\biggr\},
\end{eqnarray}


\noindent where indexes 1 and 2 at the $S$- and $D$-state of the
deuteron, denoted as $U$ and $W$, respectively, means that they
are evaluated at modulus of the proton and neutron 3-momentum in
the deuteron, ${\bf k}_1$ and ${\bf k}_2$. These are {\it
formally} related to the BS form factors  as
\begin{eqnarray}
U=\frac{\sqrt{\pi m}}{2E_{\bf k}-M_d}g_0(0,|{\bf k}|;s=M_d^2),
\quad
W=\frac{\sqrt{\pi m}}{2E_{\bf k}-M_d}g_2(0,|{\bf k}|;s=M_d^2).
\nonumber\end{eqnarray}

%

\section{Concluding remarks}
\label{sec:remarks}

The objective of the present study is evaluate the various relativistic
contributions to the angular distributions in deuteron photodisintegration
process. That can be done in a variety of theoretical frameworks and
dynamics. In this paper we have applied the fully relativistic formalism,
based on the Bethe--Salpeter equation for the 2N scattering amplitude and
deuteron bound state. Beyond the choice of the theoretical framework, which
is manifestly covariant at every step of the calculation, the important issue
is dynamical model of the nucleon-nucleon interaction. Pursuing the aim to
obtain clear understanding and conduct straightforward comparison with the
non-relativistic and minimally relativistic approaches, we employ the
effective separable interactions in construction the solvable dynamical model
of the deuteron.

In order to obtain ultimate results in analytical form we discard channels
containing negative-energy states in the Bethe--Salpeter amplitude of the
deuteron. We also do not include the two-body contributions to the
electromagnetic current operator and neglect the final state interaction
in the outgoing 2N state. The last two limitations constitutes the
so-called plane wave one-body approximation. The negative-energy states,
or P-states, in the deuteron vertex function are not presumably
irrelevant.  Their contributions  are expected to be significant within
the considered interval of the photon energies. In this respect our
present study is primarily of a comparative character.

Despite to these specifications, the strongest advantage in our
investigation concerns the fully covariant and rigorous description of the
bound state and the deuteron electromagnetic current. The present approach
accounts for wealth of the relativistic effects to the differential cross
section of deuteron photodisintegration: the role of relativity of the
transition matrix elements between nuclear states, the influence of the
retardation in the deuteron vertex function and one-particle propagator
and changes in amplitudes due to the Lorentz deformation and spin
precession.

Incorporation of these relativistic effects can play a crucial role in
theoretical analysis of deuteron photodisintegration even at intermediate
laboratory photon energies for the forward and backward scattering. Here
the most important contributions comes from the boost in the arguments of
the initial state vertex function and the boost on the relative energy in
the one-particle propagator due to recoil. As one is concerned with the
covariant approaches, the equal time approximation is more or less
reasonable approach from a pragmatic point of view.

Further, we can draw the following conclusions out of the present
investigation:  1) the Bethe--Salpeter approach allows one to take into
account Lorentz invariance and relativistic dynamical structure of the
two-nucleon system in the most general form. 2) The novel feature brought by
the Bethe--Salpeter approach is the retardation due to the dependence of the
bound state amplitude on the relative energy of the nucleons. In the plane
wave one-body approximation, the scattering amplitude bears the explicit
dependence on this variable and its magnitude is measured by the photon
energy. In our opinion, this turns out to be by far the most important fact
enabling to study recoil effects due to energy transfer to a nucleon by a
photon. 3) The role of the boost transformation of the spin degrees of
freedom becomes noticeable in increasing order of the photon energy at
forward scattering.

Finally, the region of high photon energies (above $E_\gamma$=500 MeV)
calls for a more complete investigation. In this energy region one needs
to construct a realistic interaction kernel in the Bethe--Salpeter
equation. Moreover, extending the above calculations includes
contributions due to P-states and the two-body processes in the EM current
operator matrix elements.

\acknowledgments

We are indebted to V.V. Burov, S.G. Bondarenko, A.A. Goy, A.V.  Molochkov
and A.V. Shebeko for many useful discussions and encouragement. This work
has been partially supported by grant 02.01.22/2000, ``Universities of
Russia".

\newpage

\centerline{FIGURE CAPTION}

\vspace{1cm}

FIG. 1. Diagrams corresponding to the plane wave one-body
 approximation to
the matrix element of deuteron photodisintegration.  Outgoing particles are
on their mass-shell.

\vspace{1.0cm}

FIG. 2. Modulus of the energy $|k_0|$ and 3-momentum $|{\bf k}|$
of a nucleon in the deuteron in its rest frame versus the photon
energy $E_\gamma$ (solid lines). Dash lines correspond to
solid ones while neglecting the boost parameter, i.~e.
$\sqrt{\eta}=0$ (static approximation). A set of fixed proton
scattering angles $\Theta_p=0^\circ$, 90 and 180$^\circ$ in the c.
m. frame correspond to curves labeled as 1, 2 and 3, respectively.
The same curves are related to the energy and 3-momentum of a
spectator nucleon at c.m. frame angles $\Theta_p=180^\circ$, 90
and 0$^\circ$.

\vspace{1.0cm}

FIG. 3. The differential cross section in the plane wave one-body
approximation at different photon energies $E_\gamma$. Curves corresponds
to different probabilities of $^3D_1^{+}$ partial state. Solid line, $P_2
= 4$~\%, dash line, $P_2 = 6$~\%.

\vspace{1.0cm}

FIG. 4. The differential cross section in the plane wave one-body
approximation at different photon energies $E_\gamma$. Curves: solid line ---
the exact positive-energy BS calculation, dotted line --- the static
approximation (exclusion of the Lorentz contraction), dash line --- the
static approximation with taking into account the boost on one-particle
propagator due to recoil (the Lorentz contraction), 3-dot-dash line --- the
contribution due to the spin precession, Eqn.~(\ref{spin-boost}).
Probability of $^3D_1^{+}$ partial state is $P_2 = 4$~\%.

\vspace{1.0cm}

FIG. 5. The relative deviation of the deuteron photodisintegration cross
section for the photon energies $E_\gamma$ = 100, 300 and 500 MeV for the
approximations with respect to the BS result: BS-ZO (dash line) and BS-BR
(dot-dash line). On the Y-axis it is plotted
$\case{\sigma_0-\sigma_{\text{BS-}\alpha}}{\sigma_0}\times 100~\%$, where
$\alpha=$BR, ZO. Probability of $^3D_1^{+}$ partial state is $P_2 = 4$~\%.

\vspace{1.0cm}

FIG. 6. The relative difference of the deuteron photodisintegration cross
section for the photon energies $E_\gamma$ = 100, 300 and 500 MeV for the
following cases: the equal time approximation (dash line), the minimally
relativistic approach (dot-dash line) and non-relativistic approaches (dotted
line) with respect to the exact positive-energy BS result. On the Y-axis it
is plotted $\case{\sigma-\sigma_{\alpha}}{\sigma}\times 100~\%$, where
$\alpha=$ET, ET-SA and NR. Probability of $^3D_1^{+}$ partial state is $P_2 =
4$~\%.

\vspace*{2cm}


\end{document}